\documentclass[twocolumn,superscriptaddress,amsmath, amssymb, amsfonts,preprintnumbers,aps,prd,longbibliography,nofootinbib]{revtex4-1}

\usepackage{scalerel}
\usepackage{xcolor}
\usepackage{amsmath,amsfonts,amssymb}%,calrsfs}
\usepackage[small,bf,hang]{caption}
\usepackage{slashed}
\usepackage{latexsym,epsfig}
\usepackage{dsfont}
\usepackage{arydshln}
\usepackage{extarrows}
\usepackage{hyperref}
\usepackage{array}

\def\moth{\mathsurround=0pt}
\newdimen\zo \zo=0pt

\def\tick{\leaders\hrule height 0.5ex depth 0pt \hskip 0.5pt}
\def\upboxfill{$\moth \setbox\zo\hbox{\tick}%
  \hskip 3pt\hbox to 0pt{$\tick$\hss}\hrulefill \hbox to 7.5pt{$\tick$\hss}$}

\def\dtick{\leaders\hrule height .34pt depth 0.5ex \hskip 0.5pt}
\def\downboxfill{$\moth \setbox\zo\hbox{\dtick}%
  \hskip 2pt\hbox to 0pt{$\dtick$\hss}\hrulefill \hbox to 2pt{$\dtick$\hss}$}

\def\bec{\begin{center}}
\def\ec{\end{center}}
\def\a{\alpha}

\def\nn{\nonumber}

\def\be{\begin{equation}}
\def\ee{\end{equation}}
\def\bea{\begin{eqnarray}}
\def\eea{\end{eqnarray}}
\def\ba{\begin{array}}
\def\ea{\end{array}}

\begin{document}

\title{Trivialization of the gravitational Green-Schwarz transformation in the non-relativistic limit of string theory}

\author{Eric Lescano} 
\email{eric.lescano@uwr.edu.pl}
\affiliation{Institute for Theoretical Physics (IFT), University of Wroclaw, \\
pl. Maxa Borna 9, 50-204 Wroclaw,
Poland}

\begin{abstract}
We show that the gravitational Green–Schwarz (GS) transformation becomes trivial in the non-relativistic (NR) limit of ten-dimensional heterotic supergravity with four-derivative corrections. This constitutes an important step towards establishing the trivialization of the GS mechanism in this limit. In this work, we perform a NR expansion of the Kalb–Ramond field and identify the finite Green–Schwarz transformation in this limit, which can be interpreted as a non-covariant $SO(8)$ transformation. We then construct an explicit field redefinition such that the redefined two-form is invariant under this symmetry. This result is compared with the previously reported trivialization of the gauge GS mechanism under the same limit. Both field redefinitions can be implemented simultaneously, and the associated Chern–Simons terms are exact, arising directly from the redefinition structure, and leading to a trivial Bianchi identity. These results support the expectation that anomaly cancellation becomes automatic in the NR regime, and therefore we discuss their potential implications.

\end{abstract}

\maketitle

\section{Introduction}

Non-relativistic (NR) string theories \cite{NR1}-\cite{NR2} emerged as a rich and consistent extension of string theory, motivated by decoupling limits, dualities, and non-Lorentzian geometries (see \cite{Review} for a modern review). These theories admit worldsheet formulations with modified symmetries and target spaces governed by generalized Newton–Cartan geometry. In particular, the NR limits of supergravity provide a natural low-energy arena in which the interplay between background geometry, field content, and anomaly cancellation can be studied beyond the relativistic regime. Anomaly cancellation is a cornerstone of string theory consistency. In heterotic string theory, gauge and gravitational anomalies are canceled via the Green-Schwarz mechanism~\cite{GS}, which introduces a nontrivial transformation of the Kalb-Ramond $B$-field under both gauge and local Lorentz transformations. These transformations are essential for consistency and cannot be removed by field redefinitions in a relativistic setting. The structure of the anomaly polynomial imposes rigid constraints, most notably restricting the gauge group to either $\mathrm{SO}(32)$ or $E_8 \times E_8$, due to the requirement that the total anomaly factorizes as $I_{12}, \sim (\text{tr}F^2 - \text{tr}R^2)^2$, a distinctive feature of the heterotic string \cite{Gross}-\cite{Gross2}.  

While a previous study \cite{Lescano2025} analyzed the gauge Green–Schwarz mechanism in the NR limit—focusing on the behavior of the gauge sector and its associated two-derivative action, the present Letter addresses a different and independent problem: the gravitational Green–Schwarz mechanism and its four-derivative $\alpha'$-corrections. The two mechanisms involve distinct symmetries and Chern–Simons structures: the gauge mechanism concerns the non-Abelian field strength and its coupling to the $B$-field, while the gravitational mechanism involves the Lorentz connection and curvature. 

In this work, we demonstrate that the gravitational Green-Schwarz transformation of the $b$-field can be entirely removed by field redefinitions. As a consequence, the redefined two-form $\tilde{b}_{\mu\nu}$ becomes manifestly invariant under $SO(8)$ transformations. Moreover, the modified three-form field strength $\bar{h}_{\mu\nu\rho}$ becomes exact, and its Bianchi identity is trivially satisfied. These results, together with the previously reported gauge Green-Schwarz trivialization \cite{Lescano2025}, have profound implications: the anomaly cancellation in heterotic string theory may no longer impose constraints on the gauge group or background topology. The need for anomaly inflow could be eliminated, and the associated cohomological conditions would no longer be required. While in this work we pave the way towards the full trivialization of the Green-Schwarz mechanism, a complete cancellation would lead to a dramatic relaxation of consistency constraints and allows for compactifications and gauge sectors that would be anomalous in the relativistic theory. Similarly, the Wald entropy and the first law of thermodynamics are simplified under this limit, thanks to the simplification of the Bianchi identity for the three-form curvature. These findings open the door to a broader landscape of consistent NR heterotic theories in the long term, including those with unconventional gauge groups, torsional geometries, or non-Kähler compactifications upon considering the NR limit. Furthermore, if the full Green-Schwarz mechanism would be trivial under this limit, this could provide a new perspective on the role of anomalies in string theory, suggesting that non-Lorentzian limits can fundamentally alter their geometric and topological origin. The four-derivative corrections to the universal action principle of the different formulations of heterotic string theory were historically computed considering three- and four-point scattering amplitudes for the massless states \cite{GrossSloan}-\cite{CN} . This method is based on the study of the different types of string interactions through the S-matrix to construct an effective Lagrangian, originally computed by Metsaev and Tseytlin\cite{MetsaevTseytlin},
\bea
S_{MT} = \int d^Dx \sqrt{-g} e^{-2\phi} (L^{(0)} + L^{(1)}_{MT}) \, ,
\label{fullA}
\eea
where $L^{(0)} = R + 4 \partial_\mu \phi \partial^{\mu}\phi - \frac{1}{12} H_{\mu \nu \lambda} H^{\mu \nu \lambda}$
and
\bea
L^{(1)}_{\rm MT} & = & - \frac{a+b}{8} \Big[  R_{\mu \nu \rho \sigma} R^{\mu \nu \rho \sigma} - \frac12 H^{\mu \nu \rho} H_{\mu \sigma \lambda} R_{\nu \rho}{}^{\sigma \lambda} + \frac{1}{24} H^4 \nn \\ &&
- \frac18 H^2_{\mu \nu} H^{2 \mu \nu} \Big] + \frac{a-b}{4} H^{\mu \nu \rho} C_{\mu \nu \rho} \, 
\label{MT}
\eea
corresponds to the first-order $\alpha'$-correction with $a$ and $b$ units of $\alpha'$. Our compact notation is
\bea
\label{CSdef}
C_{\mu \nu \rho} & = & w_{[\mu}{}^{AB} \partial_{\nu} w_{\rho]AB} + \frac23 w_{[\mu}{}^{AB} w_{\nu B}{}^{C} w_{\rho] CA} \, , \\
H^2_{\mu \nu} & = & H_{\mu}{}^{\rho \sigma} H_{\nu \rho \sigma} \, , \quad H^2  = H_{\mu \nu \rho} H^{\mu \nu \rho} \, ,
\eea
with $A,B=1,\dots,10$ a flat index and $w_{\mu AB}$ the spin connection given by
\bea
w_{\mu B C}&=&E_\mu{}^A\left(-E^\rho{}_{[A}{} E^\nu{}_{B]}{}\partial_\rho E_{\nu C} + E^\rho{}_{[A}{} E^\nu{}_{C]}{}\partial_\rho E_{\nu B} \right.
\nn \\ && \left. + E^{\rho}{}_{[B}  E^{\nu}{}_{C]} \partial_{\rho} E_{\nu A}\right)\, .
\eea

The bi-parametric form of the Lagrangian (\ref{MT}) was introduced in \cite{Tduality}. In order to find heterotic $\a'$-corrections we impose $(a,b) = (-\alpha',0)$. For the moment, we are turning off the non-Abelian gauge field $A_{\mu}{}^{i}$ of the heterotic supergravity, which we will include later. 

The full action (\ref{fullA}) requires a modification of the Lorentz transformation of the B-field at order $\alpha'$,
\bea
\delta_{\Lambda} B_{\mu \nu} = - \frac{\alpha'}{2}  \partial_{[\mu} \Lambda^{AB} w_{\nu] AB} \, .
\eea
This expression is the so-called Green-Schwarz transformation \cite{GS} and it is the cornerstone of string theory phenomenology. We can absorb the second line of (\ref{MT}) by deforming the curvature of the B-field
\bea
\bar H_{\mu \nu \rho} = 3 \Big(\partial_{[\mu} B_{\nu \rho]} + \frac{\alpha'}{2} C_{\mu \nu \rho}\Big) \, .
\eea
The new terms in the curvature of the B-field guarantees that $\bar H_{\mu \nu \rho}$ transforms trivially under Lorentz transformations and then $L_{\rm MT}$ is Lorentz invariant when four-derivative corrections are taken into account.

\section{Main Result: Trivialization of the gravitational Green-Schwarz Mechanism}
\subsection{$\alpha'$-rescaling for the non-relativistic limit}
Higher-derivative corrections in heterotic string theory arise from quantum consistency of the worldsheet sigma model, with \(\alpha'\) serving as the loop-counting parameter in target-space curvature corrections. In the NR limit, the expansion of the worldsheet action around a NR background geometry implies a systematic suppression of gravitational curvature effects. From the sigma model perspective, the worldsheet metric couples to a target-space metric with the expansion
\begin{equation}
g_{\mu\nu} = c^2 \tau_\mu^a \tau_\nu^b \eta_{ab} + h_{\mu\nu},
\end{equation}
where $a=0,1$ is a longitudinal flat index and therefore the dominant $c^2$ scaling is carried by the longitudinal Newton-Cartan clock form $\tau_\mu^a$. At leading order, the divergences of the metric are cancelled by expanding the B-field as
\bea
B_{\mu \nu} = - c^2 {\tau_\mu}^a {\tau_\nu}^{b}  \epsilon_{ab} + b_{\mu \nu} \, .
\eea
As a result, the pullback of the Riemann tensor to the worldsheet includes components that scale as $R \sim c^2$, and therefore a curvature-squared term in the effective action diverges in the $c \rightarrow \infty$ limit (this can also be checked by a straightforward computation of the Riemann squared term). To maintain a finite worldsheet theory that yields a consistent NR target-space action, we rescale the relativistic $\alpha'$ parameter in the following way, 
\begin{equation}
\alpha' \quad \longrightarrow \quad \frac{\alpha'_{NR}}{c^2},
\end{equation}
which precisely compensates the leading divergence in the two-loop beta function for the background fields. This prescription ensures that curvature-squared terms, such as those arising in the gravitational Green-Schwarz counterterm or in the $H^4$-terms, survive the NR limit up to covariant field redefinitions (see \cite{HSZ} for a discussion on the role of the covariant field redefinitions before taking the NR limit when four-derivative terms are considered). 

\subsection{Field redefinitions and Green-Schwarz trivialization}
We start by considering the NR expansion of ten-dimensional heterotic supergravity, following the formalism developed in \cite{NSNS}, where the components of the vielbein are  
\bea
E_{\mu}{}^{A} = (c \, \tau_{\mu}{}^{a}, e_{\mu}{}^{a'})
\eea
and the inverse components are
\bea
E^{\mu}{}_{A} = (\frac{1}{c} \, \tau^{\mu}{}_{a}, e^{\mu}{}_{a'}) \, .
\eea
The gravitational Green-Schwarz mechanism for the b-field is given by
\bea
\delta_{\Lambda} b_{\mu \nu} = - \frac{\alpha'_{NR}}{2 c^2} \partial_{[\mu} \Lambda^{A B} w_{\nu] AB}
\label{NRGS}
\eea
where the Lorentz parameter splits according to a SO(1,1) parameter $\Lambda_{a b} = \lambda_{M} \epsilon_{a b}$, a $SO(8)$ parameter $\Lambda_{a' b'} = \lambda_{a' b'}$ and a boost parameter $\Lambda_{a b'}= - \Lambda_{b' a} = \frac{1}{c} \lambda_{a b'}$. 

The only contribution upon considering the NR limit $c\rightarrow \infty$ is given by a transverse Green-Schwarz mechanism of the form,
\bea
\delta_{\lambda} b_{\mu \nu} = - \frac{\alpha'_{NR}}{4} \partial_{[\mu} \lambda^{a' b'} \tau_{\nu]}{}^{d} \eta_{d e} \tau_{\rho \sigma}{}^{e}  e^{\rho}{}_{a'} e^{\sigma}{}_{b'} \, ,
\eea
where $\tau_{\rho \sigma}{}^{e}=2\partial_{[\rho} \tau_{\sigma]}{}^{e}$ is the intrinsic torsion, and we have simplified the last expression by using the Newton-Cartan relations,
\begin{align}
  \tau_{\mu}{}^{a} e^{\mu}{}_{a'} & = \tau^{\mu}{}_{a} e_{\mu}{}^{a'} = 0 \, , \qquad  e_{\mu}{}^{a'} e^{\mu}{}_{b'} = \delta^{a'}_{b'} \, , \\
  \tau_{\mu}{}^{a} \tau^{\mu b} & = \eta^{a b} \, , \qquad \tau_{\mu}{}^{a} \tau^{\nu}{}_{a} + e_{\mu}{}^{a'} e^{\nu}{}_{a'} = \delta_{\mu}^{\nu} \, .
\label{NCrelations}
\end{align} 
The transversal Green-Schwarz mechanism (\ref{NRGS}) can be trivialized by the simple observation that $S_{\nu a' b'} = \tau_{\rho \sigma}{}^{e} \tau_{\nu}{}^{d} \eta_{d e} e^{\rho}{}_{a'} e^{\sigma}{}_{b'}$ transforms covariantly under $SO(8)$, i.e.,
\bea
\delta S_{\nu a' b'} = S_{\nu c' b'} \lambda^{c'}{}_{a'} + S_{\nu a' c'} \lambda^{c'}{}_{b'}\, .
\eea
The field redefinition that trivializes this mechanism is given by
\begin{equation}
\tilde{b}_{\mu\nu} = b_{\mu\nu} - \frac{\alpha'_{NR}}{4} w_{[\mu}{}^{a' b'} \tau_{\nu]}{}^{d} \eta_{d e} \tau_{\rho \sigma}{}^{e}  e^{\rho}{}_{a'} e^{\sigma}{}_{b'} \, ,
\label{fieldredefGSgrav}
\end{equation}
which exactly absorbs the non-covariant $SO(8)$ transformation of the b-field. This field redefinition leads to a finite, gauge-invariant three-form curvature $\bar{h}_{\mu\nu\rho}$ which we can define as 
\bea
\bar{h}_{\mu \nu \rho} = 3 \partial_{[\mu} \tilde b_{\nu \rho]} = h_{\mu \nu \rho} + C_{\mu \nu \rho} \, ,
\eea
where the Chern-Simons 3-form is given by
\bea
C_{\mu \nu \rho} = - \frac{\alpha'_{NR}}{4} \partial_{[\mu} \Big(w_{\nu}{}^{a' b'} \tau_{\rho]}{}^{d} \eta_{d e} e^{\epsilon}{}_{a'} e^{\sigma}{}_{b'} \tau_{\epsilon \sigma}{}^{e} \Big) \, .
\eea
\subsection{Comparison with the gauge Green-Schwarz trivialization}
For comparing the previous trivialization with the gauge one, let us introduce the gauge field $A_{\mu}{}^{i}$ with expansion $\hat{A}_{\mu}^{i} = c \ {\tau_\mu}^{-} \alpha_{-}^i + \frac{1}{c} a_\mu^i$. In this expression, $a_\mu^i$ is a gauge connection and $\alpha_{-}^i$ is a vector. This expansion was analyzed in \cite{Lescano2025}, where the trivialization of the gauge Green-Schwarz mechanism was observed \footnote{Another possibility to explore the NR limit of heterotic supergravity is to follow the prescription given in \cite{BR}. In this case the cancellation of the gauge Green-Schwarz mechanism is not straightforward due to non-covariant gauge transformations in the frame fields. However, since that prescription is T-duality invariant, there should exist field redefinitions connecting both \cite{Lescano2025} and \cite{BR}.}. Upon taking the limit $c\rightarrow \infty$ the b-field transforms as, 
\bea
\delta_{ \lambda} b_{\mu\nu} & = & -(\partial_{[\mu} \lambda^i) \alpha_{-i} \tau_{\nu]}^-\, , \label{gaugetransb} \, .
\eea
with $\lambda^i$ an arbitrary gauge parameter, while the gauge field transform as
\bea
\delta_{\lambda} a_\mu^i & = & \partial_\mu \lambda^i + f^i{}_{jk} \lambda^j a_\mu^k\, , \label{gaugetransa} \\
\delta_{\lambda} \alpha_{-}^i & = & f^i{}_{jk} \lambda^j \alpha_{-}^k\, . \label{gaugetransalpha}
\eea
Therefore, one can redefine the b-field as 
\bea
\tilde b_{\mu\nu} = b_{\mu\nu} + a_{[\mu| i} \alpha_{-}^i \tau_{|\nu] -}
\label{tildeb}
\eea
which is gauge invariant, i.e. $\delta_{\lambda}\tilde b_{\mu \nu}=0$. Since all the terms are $SO(8)$ invariant in the previous expression, both Green-Schwarz mechanisms can be trivialized simultaneously. While the leading order of the gauge Lagrangian was explicitly computed in \cite{Lescano2025}, the field redefinition (\ref{fieldredefGSgrav}) does not guarantee the finiteness of the $L^{1}$ Lagrangian, since covariant field redefinitions may still be necessary for the compensation of potential divergences coming from the Riem$^2$ term. 
\section{Implications}
A full trivialization of the gravitational and gauge Green-Schwarz mechanism in the NR limit of heterotic supergravity would lead to a profound restructuring of the constraints of heterotic string theory. Motivated by the initial results in this paper, where we trivialize the gravitational Green-Schwarz transformation, we now outline the main consequences of the full trivialization of the Green-Schwarz mechanism.
\subsection{Anomaly cancellation}
In relativistic heterotic string theory, one-loop gauge and gravitational anomalies are canceled by a nontrivial transformation of the Kalb-Ramond $B$-field, via its coupling to Chern-Simons terms \cite{GS}. In the NR limit, the Green-Schwarz transformation of the b-field is absorbed through a field redefinition, as showed in the previous section, and therefore the redefined  $\tilde{b}_{\mu\nu}$ field is invariant under both local Lorentz and gauge transformations. Consequently, it is expected that all the heterotic anomalies could be eliminated through field redefinitions and no longer appear in the effective action \footnote{The analysis of the anomalies still requires the implementation of covariant field redefinitions, which are not straightforward to implement. The best strategy is to study the action at higher-derivative orders, and guide the cancellation with the field redefinitions used to make the action finite. The author thanks his anonymous referee for this observation.}. The trivialization of this work therefore suggests that anomaly cancellation may no longer be an external consistency condition, in the NR limit of heterotic string theory. In particular, it could potentially imply that the anomaly polynomial becomes cohomologically trivial at this order.   

\subsection{Relaxation of gauge group constraints}

Relativistic heterotic string theory admits only specific gauge groups—namely $\mathrm{SO}(32)$ and $E_8 \times E_8$—as required by the factorization of the anomaly polynomial,
\begin{equation}
I_{12} \sim \left( \mathrm{tr}F^2 - \mathrm{tr}R^2 \right)^2.
\end{equation}
This condition arises from the Green-Schwarz mechanism. In contrast, the NR limit trivializes this mechanism, eliminating the need for such factorization, and potentially removing constraints on the gauge group. While the analysis in this Letter is restricted to the bosonic sector and valid up to four-derivative order in the absence of fermions, it strongly suggests that consistent NR gauge sectors can extend beyond conventional options. This includes $\mathrm{SU}(N)$, products of simple or semi-simple groups, which are excluded in the relativistic setting. However, if the NR theory is understood as a strict limit of a fixed relativistic heterotic model, the gauge group must be fixed demanding compatibility with the relativistic theory (e.g., $\mathrm{SO}(32)$ or $E_8 \times E_8$), and the NR limit inherits this choice. Therefore, the relaxation of gauge group constraints should be viewed as a genuine feature of intrinsically NR theories—that is, theories not obtained as limits of relativistic string backgrounds. From a more general perspective, the structure enabling this relaxation, i.e. the presence of a Chern-Simons-like three-form in the definition of the $H$-flux, appears in a wider class of theories beyond heterotic strings, including gauged supergravities. The trivialization mechanism uncovered here may thus reflect a deeper property of Chern-Simons couplings under NR limits \footnote{We thank to D. Marques for sharing this enlighting observation.}.

\subsection{Modified topological constraints in compactifications}

In relativistic string compactifications, anomaly cancellation imposes global topological constraints on the spacetime and gauge bundles. This is encoded in the Bianchi identity for the $H$-flux,
\begin{equation}
dH = \mathrm{tr}(F \wedge F) - \mathrm{tr}(R \wedge R),
\end{equation}
which enforces a matching of second Chern classes~\cite{GSW2}:
\begin{equation}
c_2(V) = c_2(TX).
\end{equation}
In the NR setting, the Green-Schwarz transformations are absorbed by a field redefinition of the Kalb-Ramond field: $\tilde{b} = b + D$, where $D$ is the field redefinition which trivializes the gauge and gravitational GS mechanisms, simultaneously. Crucially, the corresponding three-form field strength becomes exact:
\begin{equation}
\bar{h} = d\tilde{b} = db + dD = d(b + D),
\end{equation}
so its Bianchi identity satisfies
\begin{equation}
d\bar{h} = 0
\label{newBianchi}
\end{equation}
identically, without requiring a nontrivial cancellation between $\mathrm{tr}(F \wedge F)$ and $\mathrm{tr}(R \wedge R)$. As a result, the anomaly inflow condition may become structurally trivial, and no cohomological constraint on gauge or tangent bundles may remain. The results of this paper, together with a complete prove of the Green-Schwarz mechanism, open up the possibility of previously forbidden compactifications including those with non-matching Chern classes, non-Kähler internal geometries, or background torsion in the NR limit.

\subsection{Wald entropy and the first law in non-relativistic heterotic backgrounds}

The analysis of black hole thermodynamics in the presence of higher-derivative corrections has traditionally relied on Wald's formalism \cite{Wald}. For heterotic supergravity at order $\alpha'$, the relevant Noether charges require careful treatment due to the Green-Schwarz couplings between the Kalb-Ramond field and the gauge and gravitational Chern-Simons terms\cite{Ortin}. In the NR limit, the results of this letter predict a dramatic simplification of this structure. Since both the gauge and Lorentz Green-Schwarz mechanisms are trivialized, the redefined Kalb-Ramond field transforms covariantly, and no Chern-Simons modifications are induced. As a result, the Bianchi identity for the corrected three-form curvature reduces to a standard closure condition in the NR limit, as in Eq.~(\ref{newBianchi}). This implies that the horizontal Kalb-Ramond momentum map \( P_k \) becomes simpler, relaxing its role in the entropy computation (with a similar argument we expect a relaxation in the vertical Yang-Mills momentum map). The trivialization of the Green-Schwarz transformation in the NR limit of heterotic string theory thus enables a simpler derivation of the first law of thermodynamics, in which all thermodynamic potentials and conserved charges are computed from manifestly gauge- and Lorentz-invariant expressions, up to four-derivative contributions.

\subsection{Connection to duality-invariant frameworks}
The NR limit of heterotic string theory aligns naturally with duality-invariant formulations such as Double Field Theory (DFT) \cite{DFT1}-\cite{DFT2}, where generalized geometry encodes both geometric and gauge degrees of freedom \cite{EandD}. The four-derivative extension of DFT \cite{Tduality} \cite{Hohm} involves a generalized vielbein formalism \cite{DFTframe}. It is therefore not clear whether the NR limit produces a finite four-derivative Lagrangian considering the flux formulation of DFT, compatible with a generalization of the Green-Schwarz mechanism \cite{Odd}. This motivates a complete NR extension of heterotic DFT, incorporating supersymmetry and $\alpha'$ corrections beyond leading order, extending the results of \cite{LNR} in the NR limit.

\section{Conclusion}

We have proved that the gravitational Green-Schwarz transformation of heterotic supergravity can be trivialized in the NR limit up to four-derivative terms. We constructed the field redefinition of the b-field that makes this possible and, by construction, the Bianchi identity for the 3-form curvature does not include $\alpha'$-corrections. Our results open the door to a new paradigm for anomaly cancellation in the NR limit of heterotic string theory with several implications if the full Green-Schwarz mechanism can be proved trivial. The immediate continuation of this work includes:
\begin{itemize}
    \item Studying the role of supersymmetry when higher-derivative terms are considered, extending the seminal work in \cite{Bsusy} and exploring further trivializations of the supersymmetric transformations at the NR supergravity level \cite{Bdroo}.
    
    \item Understanding the embedding of the duality invariant construction of \cite{LNR} for the higher-derivative structure of heterotic DFT in the NR limit, including the  construction of the full four-derivative heterotic finite action (including fermions) in the NR limit. Initial progress was already done in \cite{LescanoExtra} using the Bergshoeff-de Roo identification.
    
    \item Constructing the simplified first law and Wald entropy formula of NR heterotic stringy black holes at first order in $\alpha'$. 

    \item Extending the analysis to six- and eight-derivatives order, towards a complete proof of the trivialization of the Green-Schwarz mechanism under the NR limit.
\end{itemize}
The previous ideas follow a clear path towards the understanding of the $\alpha'$- corrected heterotic supergravity under the NR limit.
\begin{widetext}
\acknowledgments
E.L. is very grateful to Diego Marques for discussions. This work is supported by the SONATA BIS grant 2021/42/E/ST2/00304 from the National Science Centre (NCN), Poland.

\bibliographystyle{apsrev4-2}

\end{widetext}
\end{document}